\documentstyle[sprocl,epsf,psfig]{article}

\def\Journal#1#2#3#4{{#1} {\bf #2}, #3 (#4)}
\def\PRD{{\em Phys. Rev.} D}

\newcommand{\vrh}{\hat{\vr}}

\newcommand{\gm}{\gamma}
\newcommand{\dl}{\delta}
\newcommand{\ep}{\epsilon}

\newcommand{\kp}{\kappa}
\newcommand{\lm}{\lambda}
\newcommand{\rh}{\rho}

\newcommand{\ph}{\phi}
\newcommand{\vr}{\varphi}

\newcommand{\Sg}{\Sigma}

\newcommand{\half}{\frac{1}{2}}

\newcommand{\eela}[1]{\label{#1}\end{equation}}
\newcommand{\eeala}[1]{\label{#1}\end{eqnarray}}
\newcommand{\be}{\begin{equation}}
\newcommand{\ee}{\end{equation}}
\newcommand{\bea}{\begin{eqnarray}}
\newcommand{\eea}{\end{eqnarray}}

\begin{document}
\title{Twin Peaks\footnote{Presented by J.~Smit.}}

\author{Mischa Sall\'e, Jan Smit and Jeroen C. Vink}

\address{Institute for Theoretical Physics, University of Amsterdam,
Valckenierstraat 65, 1018 XE Amsterdam, the Netherlands}

\maketitle\abstracts{
The on-shell imaginary part of the retarded selfenergy of massive 
$\vr^4$ theory in 1+1 dimensions is logarithmically infrared divergent. 
This leads to a zero in the spectral function,
separating its usual bump into two. The twin peaks interfere in 
time-dependent correlation functions, which causes oscillating 
modulations on top of exponential-like decay, while the usual formulas for
the decay rate fail. We see similar modulations in 
our numerical results for a mean field correlator, using a 
Hartree ensemble approximation.
}

\noindent
In our numerical simulations of 1+1 dimensional $\vr^4$ theory
using the Hartree ensemble approximation \cite{Vi00}
we found funny modulations in a time-dependent correlation function. 
Fig.~1 shows such modulations on top of
a roughly exponential decay. 
The correlation function 
is the time average of the zero momentum mode of the mean field, 
$F_{\rm mf}(t) =\overline{\vr(t)\vr(0)} - \overline{\vr(t)}\;\overline{\vr(0)}$,
where the over-bar denotes a time average, $\overline{X(t)} =
\int_{t_1}^{t_2}dt'\, X(t+t')/(t_2-t_1)$,
taken after waiting a long time $t_1$ for the system to be in approximate
equilibrium. This equilibrium
is approximately thermal 
and $F_{\rm mf}(t)$ is analogous to the symmetric correlation function 
of the quantum field theory at finite temperature,
$
F(t) = \langle\half\{\vrh(t),\,\vrh(0)\}\rangle_{\rm conn}.
$
A natural question is now,
does $F(t)$ also have such modulations? 

The function $F(t)$ can be expressed in terms of the zero momentum
spectral function $\rh(p^0)$,
\be
F(t) = \int_{-\infty}^{\infty} \frac{dp^0}{2\pi}\, 
e^{-ip^0 t}\,
\left(\frac{1}{e^{p^0/T}-1}+\half\right)\, \rh(p^0),
\label{Fdef}
\ee
and the latter in turn in terms of the retarded selfenergy $\Sg(p^0)$,
\be
\rh(p^0)=
\frac{-2{\rm Im}\,\Sg(p^0)}{[m^2 - (p^0+i\ep)^2 + {\rm
Re}\,\Sg(p^0)]^2 + [{\rm Im}\,\Sg(p^0)]^2}.
\label{rhdef}
\ee
The selfenergy can be calculated in perturbation theory. The one and two loop
diagrams in the imaginary time formalism which have nontrivial energy-momentum
dependence are shown in Fig.~2. Diagrams not shown
give only rise to an effective temperature dependent 
mass, which we assume to be the mass in the propagators of the diagrams 
in Fig.~2, after adding a counterterm that sets the real part
of $\Sg$ to zero at $p^0 = m$.
\begin{figure}[h]
\centerline{\psfig{figure=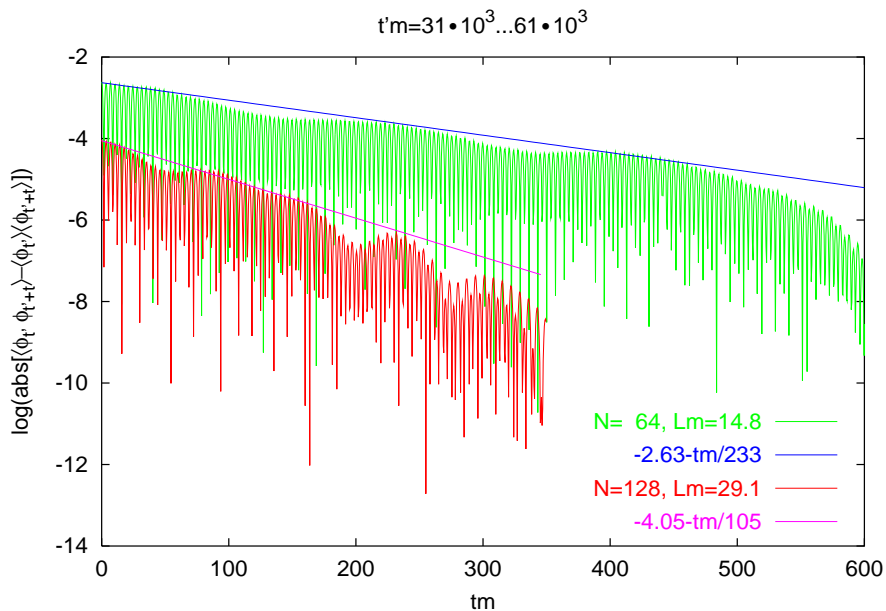,width=12.0cm}}
\caption[a]{Numerically computed correlation $\ln|F_{\rm mf}(t)|$ 
versus time $t$ in units of the inverse temperature dependent mass $m$.
The coupling is weak, $\lm/m^2 = 0.11$ and the temperature $T/m\approx 1.4$
for the smaller volume (with significant deviations from the Bose-Einstein
distribution) and $\approx 1.6$ for the larger volume (reasonable BE).
}
\vspace{-.5cm}
\end{figure}
\begin{figure}[h]
\centerline{\psfig{figure=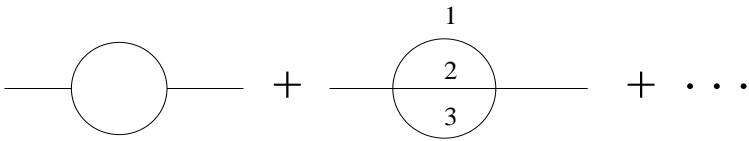,width=10.0cm}}
\caption{
Diagrams leading to thermal damping. 
}
\end{figure}
The one loop diagram is present only in the `broken phase'
(for which $\langle\vrh\rangle \neq 0$; there is really only a
symmetric phase in 1+1 dimensions, but this is due to symmetry restoration by
nonperturbative effects which will not obliterate the one-loop damping.)
The corresponding selfenergy 
has been calculated in \cite{We83}, for example. It 
only
leads to damping for frequencies $p_0^2> 4 m^2$, which are irrelevant for the
quasiparticle damping at $p_0^2 = m^2$. 
So from now on we concentrate on the two-loop diagram. 
After analytic continuation to real time one finds that it is given
by the sum of two terms, $\Sg_1 + \Sg_2$ (see e.g.\ \cite{WaHe96}).
The first 
has an imaginary part corresponding to
$1\leftrightarrow 3$ processes requiring $p_0^2 > 9m^2$, so it does not
contribute to plasmon damping.
The second is given by
\bea
\Sg_2 &=&-\frac{9\lm^2}{16\pi^2}\int 
\frac{dp_2\, dp_3}{E_1E_2E_3}\,
\frac{
(1+n_1)n_2 n_3-n_1 (1+n_2) (1+n_3)}{p^0+i\ep +E_1 - E_2 - E_3} 
\nonumber\\&&
+ \left[(p^0 +i\ep)\to-(p^0+i\ep)\right],
\label{Sg2}
\eea
where $\lm$ is the coupling constant (introduced as ${\cal L}_1 = -\lm\vr^4/4$),
and
$E_1 = \sqrt{m^2 + (p_2 + p_3)^2}$,
$E_i = \sqrt{m^2 + p_i^2}$, $i=2,3$,
$n_i = [\exp(E_i/T)-1]^{-1}$,
$i=1,2,3$.
Its imaginary part corresponds to $2\leftrightarrow 2$ processes, which
contribute in the regions near $p_0=\pm m$.

Now the usual definition of the thermal plasmon damping
rate (at zero momentum) in terms of the retarded selfenergy, 
\be
\gm = -{\rm Im}\, \Sg(m)/2m, 
\label{gmdef1}
\ee
leads to a {\em divergent} answer (a collinear divergence).
A natural way out of this difficulty may be to continue the selfenergy
analytically into the lower half of its second Riemann sheet,
$p^0\to m-i\gm$,
and replace (\ref{gmdef1}) by the improved definition
\be
m^2 - (m-i\gm)^2 + \Sg(m-i\gm) = 0.
\label{gmdef2}
\ee
The analytic continuation of the selfenergy into the region
Im $p^0 <0$
poses the puzzle how to deal with the logarithmic branch point coming
from the collinear singularity at
$p^0 = m$.
However, the ambiguity is present only in the real part of $\Sg$. 
For weak coupling $\lm/m^2\ll 1$ we get from (\ref{gmdef2}) the equation
\be
\frac{\gm}{m} = \frac{9\lm^2}{16\pi m^4}\, 
\frac{e^{m/T}}{\left(e^{m/T}-1\right)^2}\,
\left[\ln\frac{m}{\gm}+c(T)\right].
\label{gmeq}
\ee
The constant $c$ 
has to be determined by matching a numerical evaluation of
$\Sg$ to the logarithmic singularity at $p^0 = m$.

We evaluated
$\Sg_2$
in (\ref{Sg2}) for
$T=m$
by numerical integration with
$\ep/m = 0.02$, $0.01$
and linear extrapolation
$\ep\to 0$,
giving $c\approx -0.51$. 
For example, 
Eq.\ (\ref{gmeq}) now gives $\gm/m = 0.061$, 
for $\lm/m^2 = 0.4$. 

To see how well this $\gm$ 
describes the decay of the correlator $F(t)$ we evaluated this function
directly from (\ref{Fdef}) and (\ref{rhdef}).
The divergence in ${\rm Im}\,\Sg(p^0)$ at $p^0 = m$
leads to a 
{\em zero} in the spectral function $\rh(p^0)$. So is there 
a peak at all in
$\rh(p^0)$? Fig.\ 3 shows what happens: 
the `usual' peak has separated into two twins!
Fig.\ 4 shows the resulting $F(t)$. 
The effect of the double peak is indeed an oscillating modulation on top
of the roughly exponential decay. The decay corresponding to 
$\exp(-\gm t)$, with $\gm$ given by (\ref{gmeq}),
is also indicated in the plot: it
does not do a good job in describing the average decay beyond the first
interference minimum. 
The `Twin Peaks' phenomenon implies
that the usual definition of damping rate (\ref{gmdef2})
is unreliable in 1+1 dimensions.

Acknowledgements. 
We thank Gert Aarts for useful conversations. 
This work is supported by FOM/NWO.

\begin{figure}[h]
\centerline{\psfig{figure=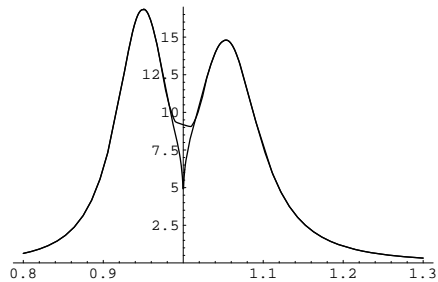,width=6.0cm}}
\caption{ 
The spectral function $\rh(p^0)$ near $p^0 = m =1$ 
corresponding to the selfenergy shown in Figs.\ 4, 5
($T=m$, $\lm=0.4 m^2$).
} 
\vspace{-.5cm}
\end{figure}
\begin{figure}[h]
\centerline{\psfig{figure=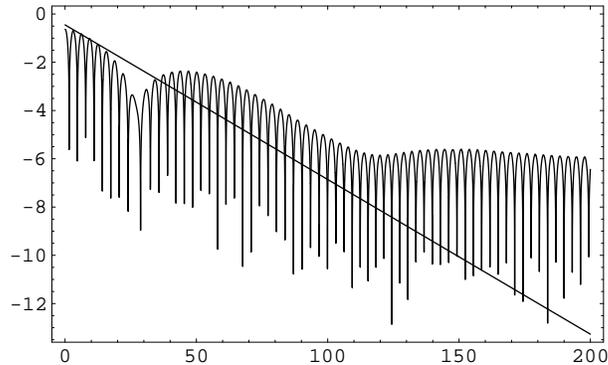,width=8.0cm}}
\caption{
Plot of $\ln|F(t)|$ versus $mt$ for $T=m$, $\lm= 0.4 m^2$.
The straight line represents $\exp(-\gm t)$.
} 
\vspace{-.5cm}
\end{figure}

\end{document}